\documentclass[aps, prd, reprint, nofootinbib]{revtex4-2}
\usepackage[utf8]{inputenc}
\usepackage[T1]{fontenc}
\usepackage{amsmath}
\usepackage{amssymb}
\usepackage{bm}
\usepackage{physics}
\usepackage{mathtools}
\usepackage{mathrsfs}
\usepackage{graphicx}
\usepackage{xcolor}
\usepackage{hyperref}

\begin{document}

\title{Gluon and Faddeev--Popov ghost propagators in rotating background: An eigenfunction approach}
\author{Benjamín García-Contreras}
\affiliation{Departamento de Física, Universidad Autónoma Metropolitana-Iztapalapa, Avenida San Rafael Atlixco 186, Ciudad de México 09340, Mexico.}
\author{Luis A. Hernández}
\affiliation{Departamento de Física, Universidad Autónoma Metropolitana-Iztapalapa, Avenida San Rafael Atlixco 186, Ciudad de México 09340, Mexico.}
\author{R. Zamora}
\affiliation{Centro de Investigaci\'on y Desarrollo en Ciencias Aeroespaciales (CIDCA), Academia Politécnica Aeronáutica, Fuerza A\'erea de Chile, Casilla 8020744, Santiago, Chile}
\affiliation{Facultad de Ingenier\'ia y Arquitectura, Universidad Central de Chile, Av. Sta. Isabel 1186, Santiago 8330601, Chile.}
\date{\today}

\begin{abstract}
We construct the free gluon and Faddeev--Popov ghost propagators in rigidly rotating cylindrical coordinates using an eigenfunction approach within the Feynman gauge. Starting from the quadratic gauge-fixed QCD action, we derive the corresponding kinetic operators and obtain their spectral representations. For the gluon field, a circular polarization basis diagonalizes the vector wave operator, leading to polarization-dependent shifts in the orders of the radial Bessel functions. In the absence of radial boundaries, we explicitly recover the standard Minkowski propagator in inertial Cartesian coordinates, showing that the off-diagonal tensor components in the cylindrical basis have a geometrical origin. We then extend the construction to a finite-radius cylinder satisfying the causality condition $|\Omega|R<1$, imposing relative boundary conditions on the gluon field and Dirichlet conditions on the ghosts. The resulting discrete transverse spectra involve both Dirichlet and Neumann eigenvalues for the gluon modes, from which we obtain explicit spectral representations of the propagators. Our results provide a consistent framework for perturbative QCD calculations in rotating geometries, retaining the angular-momentum, polarization, and boundary effects relevant to future studies of correlation functions and radiative corrections.
\end{abstract}

\maketitle

\section{Introduction \label{sec1}}

The role of angular momentum and vorticity in strongly interacting matter has become an important subject in the study of relativistic heavy-ion collisions~\cite{STAR:2017ckg,Becattini:2020ngo,Becattini:2022zvf}. The possibility of describing strongly interacting systems in global rotation has motivated theoretical investigations of how rotational motion influences the properties of quantum fields and the dynamics of their interactions~\cite{Jiang:2016woz,Fukushima:2018grm,Ayala:2020ndx,Ayala:2021osy,Braguta:2021jgn,Chen:2022smf,Chernodub:2022veq,Ayala:2023vgv,Cao:2023olg,Jiang:2024zsw,Braguta:2024zpi,Fukushima:2025hmh}. Within quantum chromodynamics (QCD), these investigations concern not only the quark degrees of freedom but also the gluon fields associated with the non-Abelian gauge structure of the strong interaction. A consistent field-theoretical description of rotating QCD systems therefore requires understanding how these fundamental degrees of freedom and their correlation functions are represented in rotating coordinates.

The propagator is a central ingredient in this description, as it determines the propagation of quantum fields and provides the starting point for perturbative calculations. Considerable progress has been made in constructing propagators for scalar and fermionic fields in rotating systems, using
proper-time methods and spectral representations~\cite{Vilenkin:1980zv,Gaspar:2023nqk,Ayala:2021osy,Hernandez:2026lyd}. These developments have supported investigations of the thermodynamic and phenomenological properties of rotating matter~\cite{Siri:2024scq,Gecic:2026iry,WeiMingHua:2020eee,
Castano-Yepes:2025zae,Kawaguchi:2025mkh}. The vector sector presents additional challenges because the field components are coupled through the geometrical structure of cylindrical coordinates. In particular, the propagation of massive vector fields under rotation has been investigated
within Proca theory, where the corresponding modes and propagators can be constructed through a spectral approach \cite{Xu:2021kjg}. For massless gauge fields, however, the presence of gauge redundancy requires a different treatment.

In QCD, the gluon field is associated with a non-Abelian gauge symmetry. Its quantization requires a gauge-fixing prescription, which introduces the Faddeev--Popov ghost fields. Although ghosts do not correspond to physical asymptotic states, their contributions are essential for the consistency of the gauge-fixed theory. Unlike in Abelian gauge theories, the Faddeev--Popov fields in QCD interact with the gauge sector and contribute to quantum corrections. Consequently, a formulation of QCD in rotating coordinates must consistently account for both gluon and ghost degrees of freedom. Even at the level of the free theory, their propagators provide the building blocks needed to organize subsequent perturbative calculations in a rotating setting.

The behavior of gauge fields in rotating reference frames has been investigated from several complementary perspectives. At the classical level, Maxwell electrodynamics can be formulated in noninertial frames using its generally covariant structure~\cite{Obukhov:2021ayi}. Studies of electromagnetic waves in rotating systems have explored their mode structure and the restrictions associated with the light cylinder~\cite{PhysRevA.66.042103}. The coupling between electromagnetic helicity and rotation has also been examined through the frequency shifts experienced by rotating observers~\cite{Hauck:2003gy,Mashhoon:2024wvp}. At the quantum level, investigations of rotating photon systems have revealed vortical transport phenomena associated with helicity and zilch currents~\cite{Avkhadiev:2017fxj,Yamamoto:2017uul,Chernodub:2018era,Huang:2020kik}. Related developments based on the photon Wigner function have connected these effects with kinetic descriptions~\cite{Huang:2020kik,Hattori:2020gqh}. More recently, thermal gauge theories in global rotation have been formulated by incorporating angular momentum into the equilibrium density operator, providing a framework for the treatment of gauge and ghost fields~\cite{Salvio:2026ewl}. Together, these studies motivate a direct construction of the free gluon and ghost propagators in rotating coordinates, with particular attention to their spectral and tensorial structures. More recently, the quantization of the electromagnetic field in rotating coordinates has been investigated in detail, including the construction of physical photon modes, their canonical quantization, and the corresponding time-ordered propagator. The full gauge-fixed photon propagator has also been obtained through complementary approaches based on the transformation of the inertial propagator and the spectral representation of the vector wave operator \cite{Ayala:2026ixw}. These developments provide an important reference for extending the analysis to the gauge sector of QCD, where the gluon and Faddeev–Popov ghost fields must be treated within a common gauge-fixed formulation.

An additional consideration arises from the geometry of rigidly rotating systems. In cylindrical coordinates, the rotating description is restricted by the light-cylinder condition, which requires $|\Omega| R<1$ for a cylinder of radius $R$. When the fields are confined to a finite cylindrical region, boundary conditions must also be specified, thereby modifying the allowed transverse modes. For gauge fields, this problem requires particular care, since the boundary conditions must be compatible with the gauge-fixed formulation. It is therefore useful to distinguish the effects associated with the rotating coordinate system from those arising from the spectral restrictions imposed by a finite spatial boundary.

In this work, we construct the free gluon and Faddeev--Popov ghost propagators in rigidly rotating cylindrical coordinates within the Feynman gauge. Starting from the quadratic gauge-fixed QCD action, we derive the corresponding differential operators and determine their eigenfunctions. For the gluon field, we introduce a circular basis that diagonalizes the vector wave operator, leading to radial eigenfunctions with angular indices shifted according to the circular polarization. These eigenfunctions are then used to construct spectral representations of the ghost and gluon propagators in the absence of radial boundaries. We explicitly reconstruct the tensor structure of the gluon propagator in an orthonormal basis and establish its relation to the standard Cartesian representation in Minkowski spacetime, thereby identifying the geometrical origin of the angular dependence.

We subsequently extend the construction to a cylinder of finite radius, subject to the causality condition $|\Omega|R<1$. We impose relative boundary conditions on the gluon field and Dirichlet conditions on the Faddeev--Popov ghosts, and determine the resulting discrete transverse spectra. In particular, the gluon modes separate into families associated with Dirichlet and Neumann eigenvalues, allowing us to construct the corresponding spectral propagator and its explicit tensor representation. The finite-cylinder ghost propagator is obtained from the Dirichlet eigenfunctions. This formulation distinguishes the geometrical effects of rotating coordinates from the modifications of the mode spectrum induced by the boundary.

The remainder of this work is organized as follows. In Sec.~\ref{sec2}, we derive the free gluon and ghost field equations in rotating cylindrical coordinates, diagonalize the gluon wave operator, and construct the corresponding eigenfunctions. In Sec.~\ref{sec3}, we obtain the spectral representations of both propagators, establish their connection with the standard Minkowski expressions, and extend the construction to a finite cylindrical domain with appropriate boundary conditions. Finally, in Sec.~\ref{sec4}, we summarize our results and present our conclusions.

\section{Gluon and ghost fields in a rotating background \label{sec2}}

In this section, we derive the equations of motion for free gluon and Faddeev--Popov ghost fields in rotating cylindrical coordinates. Starting from the quadratic QCD gauge action, we identify the corresponding differential operators and express them in a basis adapted to the rotational symmetry. We then construct their eigenfunctions, which will provide the basis for obtaining the propagators.

\subsection{Quadratic gauge-fixed QCD action \label{subsec2.1}}

We consider the gluonic sector of QCD, described by the action
\begin{equation}
S_g=-\frac{1}{4}\int d^4x\,\sqrt{-g}\, F_{\mu\nu}^a F^{a\mu\nu},
\label{eq:gluon_action}
\end{equation}
where $g$ is the determinant of the spacetime metric and
\begin{equation}
F_{\mu\nu}^a=\partial_\mu A_\nu^a-\partial_\nu A_\mu^a+g_s f^{abc}A_\mu^b A_\nu^c.
\label{eq:gluon_field_strength}
\end{equation}
Here, $A_\mu^a$ denotes the gluon field, $g_s$ is the strong coupling, and $f^{abc}$ are the structure constants of $SU(3)$, with $a=1,\ldots,8$.

Our purpose is to construct the free gluon and ghost propagators in rotating cylindrical coordinates. We therefore restrict our analysis to the quadratic part of the gauge-fixed QCD action. These terms define the free theory, while higher-order contributions describe gluon self-interactions. The field strength entering the quadratic action is then
\begin{equation}
F_{\mu\nu}^{a(1)}=\partial_\mu A_\nu^a-\partial_\nu A_\mu^a,
\label{eq:linear_field_strength}
\end{equation}
and the corresponding quadratic action is
\begin{equation}
S_g^{(2)}=-\frac{1}{4}\int d^4x\,\sqrt{-g}\,F_{\mu\nu}^{a(1)}F^{a(1)\mu\nu}.
\label{eq:quadratic_gluon_action}
\end{equation}
To construct the gluon propagator, we introduce the covariant
gauge-fixing term
\begin{equation}
S_{\mathrm{gf}}=-\frac{1}{2\xi}\int d^4x\,\sqrt{-g}\, \left(\nabla_\mu A^{a\mu}\right)^2,
\label{eq:gauge_fixing_action}
\end{equation}
where $\xi$ is the gauge parameter. The corresponding Faddeev--Popov ghost action is
\begin{equation}
S_{\mathrm{gh}}=-\int d^4x\,\sqrt{-g}\, \bar c^a \nabla_\mu D^{ab\mu}c^b,
\label{eq:ghost_action}
\end{equation}
with
\begin{equation}
D_\mu^{ab}=\delta^{ab}\nabla_\mu+g_s f^{acb}A_\mu^c.
\label{eq:adjoint_derivative}
\end{equation}
The ghost fields $c^a$ and $\bar c^a$ are Grassmann-valued spacetime scalars transforming in the adjoint representation of $SU(3)$.

At quadratic order, the gauge-fixed gluon action can be written, after integration by parts, as
\begin{equation}
S_A^{(2)}=\frac{1}{2}\int d^4x\,\sqrt{-g}\, A_\nu^a H^\nu{}_\sigma A^{a\sigma},
\label{eq:quadratic_gauge_fixed_action}
\end{equation}
where
\begin{equation}
H^\nu{}_\sigma=\delta^\nu{}_\sigma\nabla_\mu\nabla^\mu-R^\nu{}_\sigma-\left(1-\frac{1}{\xi}\right)\nabla^\nu\nabla_\sigma.
\label{eq:gluon_operator_general}
\end{equation}
Here, $R^\nu{}_\sigma$ is the Ricci tensor. Since the rotating coordinates considered below describe flat Minkowski spacetime, $R^\nu{}_\sigma=0$. Choosing the Feynman gauge, $\xi=1$, the free gluon equation becomes
\begin{equation}
\nabla_\mu\nabla^\mu A^{a\nu}=0.
\label{eq:free_gluon_equation}
\end{equation}

For the ghost fields, the quadratic part of Eq.~(\ref{eq:ghost_action}) is
\begin{equation}
S_{\mathrm{gh}}^{(2)}=-\int d^4x\,\sqrt{-g}\, \bar c^a\nabla_\mu\nabla^\mu c^a,
\label{eq:quadratic_ghost_action}
\end{equation}
which gives
\begin{equation}
\nabla_\mu\nabla^\mu c^a=0.
\label{eq:free_ghost_equation}
\end{equation}

Although Eqs.~(\ref{eq:free_gluon_equation}) and~(\ref{eq:free_ghost_equation}) have the same covariant form, their differential operators act on fields of different tensorial character. The covariant derivatives in the gluon equation act on a spacetime vector and therefore include connection terms, whereas the ghosts are spacetime scalars. This distinction becomes explicit when the equations are written in rotating cylindrical coordinates.

\subsection{Gluon and ghost field equations in rotating coordinates \label{subsec2.2}}

We introduce cylindrical coordinates $(t,\rho,\varphi,z)$ rotating uniformly around the $z$ axis with angular velocity $\Omega$. Their relation to inertial cylindrical coordinates is
\begin{equation}
t_I=t,\qquad
\rho_I=\rho,\qquad
\varphi_I=\varphi+\Omega t,\qquad
z_I=z,
\label{eq:rotating_transformation}
\end{equation}
where the subscript $I$ denotes the inertial reference frame. Starting from the Minkowski line element in cylindrical coordinates,
\begin{equation}
ds^2=dt_I^2-d\rho_I^2-\rho_I^2d\varphi_I^2-dz_I^2,
\end{equation}
we obtain
\begin{equation}
ds^2=(1-\Omega^2\rho^2)dt^2-2\Omega\rho^2dt\,d\varphi-d\rho^2-\rho^2d\varphi^2-dz^2.
\label{eq:rotating_metric}
\end{equation}
The corresponding metric and its inverse are
\begin{equation}
g_{\mu\nu}=\begin{pmatrix}
1-\Omega^2\rho^2 & 0 & -\Omega\rho^2 & 0\\
0 & -1 & 0 & 0\\
-\Omega\rho^2 & 0 & -\rho^2 & 0\\
0 & 0 & 0 & -1
\end{pmatrix},
\label{eq:metric_covariant}
\end{equation}
and
\begin{equation}
g^{\mu\nu}=\begin{pmatrix}
1 & 0 & -\Omega & 0\\
0 & -1 & 0 & 0\\
-\Omega & 0 & \Omega^2-\rho^{-2} & 0\\
0 & 0 & 0 & -1
\end{pmatrix}.
\label{eq:metric_contravariant}
\end{equation}
The metric determinant is $g=-\rho^2$, so that
$\sqrt{-g}=\rho$.

For the ghost field, which is a spacetime scalar, the covariant
d'Alembertian can be written as
\begin{equation}
\nabla_\mu\nabla^\mu c^a=\frac{1}{\sqrt{-g}}
\partial_\mu\left(\sqrt{-g}\,g^{\mu\nu}\partial_\nu c^a\right).
\label{eq:scalar_covariant_box}
\end{equation}
Using Eqs.~(\ref{eq:metric_covariant}) and~(\ref{eq:metric_contravariant}), we obtain
\begin{equation}
\Box_\Omega=(\partial_t-\Omega\partial_\varphi)^2-\partial_\rho^2-\frac{1}{\rho}\partial_\rho-\frac{1}{\rho^2}\partial_\varphi^2-\partial_z^2.
\label{eq:rotating_scalar_box}
\end{equation}
The free ghost equation therefore becomes
\begin{equation}
\Box_\Omega c^a=0.
\label{eq:rotating_ghost_equation}
\end{equation}
For the gluon field, the covariant derivatives also act on the spacetime vector index. The corresponding equation is
\begin{equation}
\nabla_\mu\nabla^\mu A^{a\nu}=0,
\label{eq:rotating_gluon_covariant}
\end{equation}
where
\begin{equation}
\nabla_\mu A^{a\nu}=\partial_\mu A^{a\nu}+\Gamma^\nu_{\mu\lambda}A^{a\lambda},
\label{eq:vector_covariant_derivative}
\end{equation}
and the Christoffel symbols are determined by
\begin{equation}
\Gamma^\lambda_{\mu\nu}=\frac{1}{2}g^{\lambda\sigma}
\left(\partial_\mu g_{\sigma\nu}+\partial_\nu g_{\sigma\mu}-\partial_\sigma g_{\mu\nu}\right).
\label{eq:christoffel_definition}
\end{equation}

For the rotating metric, the nonvanishing Christoffel symbols are
\begin{equation}
\begin{aligned}
\Gamma^\rho_{tt} &= -\Omega^2\rho,
&
\Gamma^\rho_{t\varphi} &= -\Omega\rho,
&
\Gamma^\rho_{\varphi\varphi} &= -\rho,
\\
\Gamma^\varphi_{t\rho} &= \frac{\Omega}{\rho},
&
\Gamma^\varphi_{\rho\varphi} &= \frac{1}{\rho},
\end{aligned}
\label{eq:rotating_christoffel}
\end{equation}
together with those obtained by symmetry in the lower indices.

To obtain the gluon equations explicitly, we evaluate the
covariant vector d'Alembertian,
\begin{equation}
\begin{aligned}
\nabla_\mu\nabla^\mu A^{a\nu}=g^{\mu\lambda}
\big[&\partial_\mu(\nabla_\lambda A^{a\nu})+\Gamma^\nu_{\mu\sigma}\nabla_\lambda A^{a\sigma} \\
&-\Gamma^\sigma_{\mu\lambda}\nabla_\sigma A^{a\nu}\big].
\end{aligned}
\label{eq:vector_dalembertian_explicit}
\end{equation}
Substituting the Christoffel symbols, we find
\begin{equation}
\begin{aligned}
\nabla_\mu\nabla^\mu A^{at}&=\Box_\Omega A^{at},
\\[2mm]
\nabla_\mu\nabla^\mu A^{a\rho}&=
\left(\Box_\Omega+\frac{1}{\rho^2}\right)A^{a\rho}
+\frac{2}{\rho}\partial_\varphi
\left(A^{a\varphi}+\Omega A^{at}\right),
\\[2mm]
\nabla_\mu\nabla^\mu A^{a\varphi}&=
\Box_\Omega A^{a\varphi}-\frac{2}{\rho}\partial_\rho \left(A^{a\varphi}+\Omega A^{at}\right)
-\frac{2}{\rho^3}\partial_\varphi A^{a\rho},
\\[2mm]
\nabla_\mu\nabla^\mu A^{az}&=\Box_\Omega A^{az}.
\end{aligned}
\label{eq:gluon_component_equations}
\end{equation}

The temporal and longitudinal components satisfy equations involving only the scalar operator $\Box_\Omega$. The transverse components, however, contain additional terms arising from the spacetime connection. These terms couple the radial and angular components and introduce an explicit dependence on the temporal component through the angular combination $A^{a\varphi}+\Omega A^{at}$.

The resulting equations show how rotation enters the free gluon operator. They also motivate the introduction of an orthonormal basis, in which the transverse structure can be simplified.

\subsection{Diagonalization of the gluon operator \label{subsec2.3}}

The equations obtained in the previous subsection couple the radial and angular components of the gluon field. To simplify their structure, we introduce the orthonormal basis defined by
\begin{equation}
\begin{aligned}
e^{\hat t} &= dt,& e^{\hat\rho} &= d\rho, \\
e^{\hat\varphi} &= \rho(d\varphi+\Omega dt),& e^{\hat z} &= dz.
\end{aligned}
\label{eq:orthonormal_coframe}
\end{equation}
In this basis, the line element takes the Minkowski form
\begin{equation}
ds^2=\eta_{\hat\mu\hat\nu} e^{\hat\mu}e^{\hat\nu},
\qquad
\eta_{\hat\mu\hat\nu}=\mathrm{diag}(1,-1,-1,-1).
\label{eq:orthonormal_metric}
\end{equation}
The corresponding gluon components are related to those in the coordinate basis through
\begin{equation}
\begin{aligned}
A^{a\hat t} &= A^{at}, & A^{a\hat\rho} &= A^{a\rho}, \\
A^{a\hat\varphi} &= \rho(A^{a\varphi}+\Omega A^{at}), & A^{a\hat z} &= A^{az}.
\end{aligned}
\label{eq:gluon_orthonormal_components}
\end{equation}

Using these relations, the free gluon equations can be expressed as
\begin{equation}
H^{\hat\mu}{}_{\hat\nu}A^{a\hat\nu}=0,
\label{eq:orthonormal_gluon_equation}
\end{equation}
where
\begin{equation}
H^{\hat\mu}{}_{\hat\nu}=\begin{pmatrix}
\Box_\Omega & 0 & 0 & 0\\
0 & \Box_\Omega+\rho^{-2} & 2\rho^{-2}\partial_\varphi & 0\\
0 & -2\rho^{-2}\partial_\varphi & \Box_\Omega+\rho^{-2} & 0\\
0 & 0 & 0 & \Box_\Omega
\end{pmatrix}.
\label{eq:orthonormal_gluon_operator}
\end{equation}
The temporal and longitudinal components are already decoupled. The remaining coupling involves only the two transverse components, $A^{a\hat\rho}$ and $A^{a\hat\varphi}$.

To diagonalize this sector, we introduce the circular components
\begin{equation}
A^{a\pm}=\frac{1}{\sqrt{2}}\left(A^{a\hat\rho}\pm iA^{a\hat\varphi}\right).
\label{eq:circular_gluon_components}
\end{equation}
Adding and subtracting the transverse equations, we obtain
\begin{equation}
H_+A^{a+}=0,
\qquad
H_-A^{a-}=0,
\label{eq:circular_gluon_equations}
\end{equation}
with
\begin{equation}
H_\pm=\Box_\Omega+\frac{1}{\rho^2}\mp\frac{2i}{\rho^2}\partial_\varphi.
\label{eq:circular_gluon_operators}
\end{equation}
The temporal and longitudinal components satisfy
\begin{equation}
H_0 A^{a\hat t}=0,
\qquad
H_0 A^{a\hat z}=0,
\qquad
H_0=\Box_\Omega.
\label{eq:longitudinal_temporal_operators}
\end{equation}

The gluon equations are therefore diagonal in the basis $(A^{a\hat t},A^{a+},A^{a-},A^{a\hat z})$, with the corresponding operators
\begin{equation}
H=\mathrm{diag}(H_0,H_+,H_-,H_0).
\label{eq:diagonal_gluon_operator}
\end{equation}
This representation separates the four vector components into independent differential equations. The operators $H_\pm$ contain the angular derivative associated with the circular components, while $H_0$ coincides with the scalar operator governing the free ghost field.

As shown in the following subsection, the angular structure of $H_\pm$ shifts the order of the radial Bessel functions relative to the scalar case.

\subsection{Eigenfunctions of the gluon and ghost operators \label{subsec2.4}}

We now construct the eigenfunctions of the diagonal operators obtained in the previous subsection. Since the metric coefficients in rotating cylindrical coordinates are independent of $t$, $\varphi$, and $z$, we seek simultaneous eigenfunctions of the corresponding differential operators in the separated form
\begin{equation}
\Phi_s(x)=e^{-iEt+ik_z z+il\varphi}R_s(\rho),
\label{eq:separated_eigenfunctions}
\end{equation}
where $E$ and $k_z$ are the frequency and longitudinal momentum, respectively, $l\in\mathbb{Z}$ is the angular quantum number, and $s=0,\pm1$ labels the diagonal operators.

We first consider the eigenvalue equation
\begin{equation}
H_s\Phi_s(x)=\lambda\Phi_s(x),
\label{eq:gluon_eigenvalue_equation}
\end{equation}
where
\begin{equation}
H_0=\Box_\Omega, \qquad
H_\pm=\Box_\Omega+\frac{1}{\rho^2}\mp\frac{2i}{\rho^2}\partial_\varphi.
\label{eq:unified_gluon_operators}
\end{equation}
The temporal and longitudinal gluon components, as well as the ghost field, correspond to $s=0$.

Acting on the separated eigenfunctions, the temporal and angular derivatives give
\begin{equation}
(\partial_t-\Omega\partial_\varphi)\Phi_s=-i(E+\Omega l)\Phi_s,
\label{eq:rotating_frequency}
\end{equation}
and
\begin{equation}
\partial_\varphi\Phi_s=il\Phi_s.
\label{eq:angular_eigenvalue}
\end{equation}
The combination $E+\Omega l$ corresponds to the inertial frequency. Indeed, under the coordinate transformation in Eq.~(\ref{eq:rotating_transformation}), a mode proportional to $e^{-i\omega_I t_I+il\varphi_I}$ becomes $e^{-iEt+il\varphi}$, with $E=\omega_I-\Omega l$. Thus, $E$ is the frequency conjugate to the rotating time coordinate, whereas $E+\Omega l=\omega_I$ is the inertial frequency.

Substituting Eqs.~(\ref{eq:separated_eigenfunctions}) and~(\ref{eq:unified_gluon_operators}) into Eq.~(\ref{eq:gluon_eigenvalue_equation}), we obtain
\begin{equation}
\frac{d^2R_s}{d\rho^2}+\frac{1}{\rho}\frac{dR_s}{d\rho}+\left[(E+\Omega l)^2-k_z^2+\lambda-\frac{(l+s)^2}{\rho^2}\right]R_s=0.
\label{eq:radial_eigenvalue_equation}
\end{equation}
Here, the angular terms combine according to
\begin{equation}
l^2+1+2sl=(l+s)^2, \quad s=\pm1.
\label{eq:angular_shift_identity}
\end{equation}
This relation shows that the circular components of the gluon field acquire radial equations with angular indices shifted by one unit relative to the scalar case.

Introducing the transverse momentum $k_\perp$ through
\begin{equation}
k_\perp^2=(E+\Omega l)^2-k_z^2+\lambda,
\label{eq:transverse_momentum_definition}
\end{equation}
the radial equation becomes
\begin{equation}
\left[\frac{d^2}{d\rho^2}+\frac{1}{\rho}\frac{d}{d\rho}+k_\perp^2-\frac{(l+s)^2}{\rho^2}\right]R_s(\rho)=0.
\label{eq:radial_bessel_equation}
\end{equation}
This is Bessel's differential equation of order $|l+s|$. Requiring regularity at the rotation axis, $\rho=0$, selects the regular solutions, which we choose as
\begin{equation}
R_s(\rho)=J_{l+s}(k_\perp\rho).
\label{eq:radial_bessel_solution}
\end{equation}

The corresponding eigenfunctions can therefore be written as
\begin{equation}
\Phi_s(x)=\mathcal{N}_se^{-iEt+ik_z z+il\varphi}J_{l+s}(k_\perp\rho),
\qquad s=0,\pm1,
\label{eq:gluon_ghost_eigenfunctions}
\end{equation}
with eigenvalues
\begin{equation}
\lambda=-(E+\Omega l)^2+k_z^2+k_\perp^2.
\label{eq:gluon_ghost_eigenvalues}
\end{equation}
It is worth emphasizing how the spin structure of the gluon field enters these solutions. In the circular basis, the transverse components are associated with the labels $s=\pm 1$, which account for their transformation under rotations about the $z$ axis. This information is encoded in the shifted Bessel orders $l+s$ appearing in Eq.~(\ref{eq:gluon_ghost_eigenfunctions}), whereas the eigenvalues in Eq.~(\ref{eq:gluon_ghost_eigenvalues}) are independent of $s$. Thus, the vector character of the gluon field is reflected in the structure of its eigenfunctions rather than in an explicit spin-dependent shift of the eigenvalues.

The normalization constants $\mathcal{N}_s$ are fixed by the orthogonality relations and the boundary conditions imposed on the eigenfunctions.

For $s=0$, Eq.~\eqref{eq:gluon_ghost_eigenfunctions} gives the eigenfunctions of the ghost operator and of the temporal and longitudinal gluon components. For $s=\pm1$, it gives the eigenfunctions associated with the circular transverse gluon components. Thus, the difference between the scalar
and vector sectors is encoded in the order of the radial Bessel functions.

These eigenfunctions provide the spectral basis for the construction of the free gluon and ghost propagators.

\section{Gluon and ghost propagators in a rotating cylinder \label{sec3}}

In this section, we construct the free ghost and gluon propagators using the eigenfunction expansion method. We first obtain their expressions in rotating cylindrical coordinates without imposing radial boundary conditions and establish their relation to the corresponding propagators in inertial coordinates. We then consider a cylinder of finite radius and examine the boundary conditions required to construct the propagators in the bounded system.

\subsection{Ghost propagator in rotating coordinates \label{subsec3.1}}

We begin with the Faddeev--Popov ghost propagator, defined by the time-ordered vacuum expectation value
\begin{equation}
D_{\mathrm{gh}}^{ab}(x,x')=\langle 0|T\left\{c^a(x)\bar c^b(x')\right\}|0\rangle.
\label{eq:ghost_propagator_definition}
\end{equation}
Since the ghost fields are Grassmann-valued, their time ordering is defined as
\begin{equation}
\begin{aligned}
T\left\{c^a(x)\bar c^b(x')\right\}={}&\theta(t-t')c^a(x)\bar c^b(x')\\
&-\theta(t'-t)\bar c^b(x')c^a(x).
\end{aligned}
\label{eq:ghost_time_ordering}
\end{equation}
The free ghost action, given in Eq.~(\ref{eq:quadratic_ghost_action}), can be expressed as
\begin{equation}
S_{\mathrm{gh}}^{(2)}=\int d^4x\,\sqrt{-g}\,\bar c^a K^{ab}c^b,
\qquad K^{ab}=-\delta^{ab}\Box_\Omega.
\label{eq:ghost_quadratic_operator}
\end{equation}
With this convention, the time-ordered propagator is related
to the Feynman inverse of the quadratic operator through
\begin{equation}
D_{\mathrm{gh}}^{ab}(x,x')=i(K_F^{-1})^{ab}(x,x').
\label{eq:ghost_inverse_relation}
\end{equation}
Consequently, it satisfies
\begin{equation}
\Box_\Omega D_{\mathrm{gh}}^{ab}(x,x')=-i\delta^{ab}\frac{\delta^{(4)}(x-x')}{\sqrt{-g}}.
\label{eq:ghost_green_equation}
\end{equation}

To solve this equation, we use the eigenfunctions of the scalar operator $H_0=\Box_\Omega$, obtained in
Sec.~\ref{subsec2.4}. They are
\begin{equation}
\Phi_\alpha(x)=\frac{1}{(2\pi)^{3/2}}e^{-iEt+ik_z z+il\varphi}J_l(k_\perp\rho),
\label{eq:ghost_normalized_eigenfunctions}
\end{equation}
where $\alpha=(E,k_z,k_\perp,l)$, with $E,k_z\in\mathbb{R}$, $k_\perp\geq0$, and $l\in\mathbb{Z}$. The corresponding eigenvalues are
\begin{equation}
\lambda_\alpha=-(E+\Omega l)^2+k_z^2+k_\perp^2.
\label{eq:ghost_operator_eigenvalues}
\end{equation}
These eigenfunctions satisfy the orthogonality relation
\begin{equation}
\begin{aligned}
\int d^4x\,\sqrt{-g}\,\Phi_\alpha^*(x)\Phi_{\alpha'}(x)={}&\delta(E-E')\delta(k_z-k_z')\\
&\times \delta_{ll'}\frac{\delta(k_\perp-k_\perp')}{k_\perp},
\end{aligned}
\label{eq:ghost_orthogonality}
\end{equation}
and the completeness relation
\begin{equation}
\sum_{l=-\infty}^{\infty}\int dE\,dk_z\,k_\perp dk_\perp\,\Phi_\alpha(x)\Phi_\alpha^*(x')=\frac{\delta^{(4)}(x-x')}{\sqrt{-g}}.
\label{eq:ghost_completeness}
\end{equation}
Their derivation is presented in Appendix~\ref{app:orthogonality}.

We expand the ghost propagator in this eigenfunction basis,
\begin{equation}
D_{\mathrm{gh}}^{ab}(x,x')=\delta^{ab}\sum_l\int dE\,dk_z\,k_\perp dk_\perp\,C_\alpha\Phi_\alpha(x)\Phi_\alpha^*(x').
\label{eq:ghost_propagator_expansion}
\end{equation}
Applying $\Box_\Omega$ to this expression and using the eigenvalue equation together with the completeness relation, we obtain
\begin{equation}
C_\alpha=-\frac{i}{\lambda_\alpha-i\epsilon},
\label{eq:ghost_expansion_coefficient}
\end{equation}
where $\epsilon\to0^+$ specifies the Feynman prescription. Substituting the eigenvalues and normalized eigenfunctions, the propagator becomes
\begin{equation}
\begin{aligned}
D_{\mathrm{gh}}^{ab}&(x,x')=\\{}&\frac{i\delta^{ab}}{(2\pi)^3}\sum_{l=-\infty}^{\infty}\int_{-\infty}^{\infty}dE\int_{-\infty}^{\infty}dk_z\int_0^\infty k_\perp dk_\perp\\
&\times\frac{e^{-iE(t-t')+ik_z(z-z')+il(\varphi-\varphi')}J_l(k_\perp\rho)J_l(k_\perp\rho')}{
(E+\Omega l)^2-k_z^2-k_\perp^2+i\epsilon}.
\end{aligned}
\label{eq:ghost_propagator_rotating}
\end{equation}

The dependence on the angular velocity appears through the combination $E+\Omega l$. To examine its origin, we introduce the inertial frequency $\omega_I=E+\Omega l$, so that $dE=d\omega_I$. Using the coordinate transformation in Eq.~(\ref{eq:rotating_transformation}), we find
\begin{equation}
e^{-iE(t-t')+il(\varphi-\varphi')}=e^{-i\omega_I(t_I-t_I')+il(\varphi_I-\varphi_I')}.
\label{eq:ghost_inertial_phase}
\end{equation}
The propagator can therefore be rewritten as
\begin{equation}
\begin{aligned}
D_{\mathrm{gh}}^{ab}&(x,x')=\\{}&\frac{i\delta^{ab}}{(2\pi)^3}\sum_{l=-\infty}^{\infty}\int d\omega_I\,dk_z\,k_\perp dk_\perp\\
&\times\frac{e^{-i\omega_I(t_I-t_I')+ik_z(z-z')+il(\varphi_I-\varphi_I')}J_l(k_\perp\rho)J_l(k_\perp\rho')}{\omega_I^2-k_z^2-k_\perp^2+i\epsilon}.
\end{aligned}
\label{eq:ghost_propagator_inertial}
\end{equation}
This is the cylindrical representation of the free ghost propagator in the Minkowski vacuum. Thus, in the absence of boundaries, the explicit dependence on $\Omega$ arises from the choice of rotating coordinates and does not represent a physical modification of the propagator.

To describe a system confined within a cylinder of finite radius, the eigenfunctions must also satisfy appropriate boundary conditions. These conditions determine the allowed transverse momenta and modify the spectral expansion. We postpone their implementation until Sec.~\ref{subsec3.3}, after constructing the corresponding free gluon propagator.

\subsection{Gluon propagator in rotating coordinates \label{subsec3.2}}

We now construct the free gluon propagator using the eigenfunctions obtained in Sec.~\ref{subsec2.4}. Unlike the ghost propagator, the gluon propagator carries two Lorentz indices, and its construction must account for the vector structure of the field.

The Feynman propagator is defined as the time-ordered vacuum expectation value
\begin{equation}
D^{ab\,\mu\nu}(x,x')=\langle 0|T\left\{A^{a\mu}(x)A^{b\nu}(x')\right\}|0\rangle.
\label{eq:gluon_propagator_definition}
\end{equation}
Since the gluon fields are bosonic, their time ordering is
\begin{equation}
\begin{aligned}
T\left\{A^{a\mu}(x)A^{b\nu}(x')\right\}={}&\theta(t-t')A^{a\mu}(x)A^{b\nu}(x') \\
&+\theta(t'-t)A^{b\nu}(x')A^{a\mu}(x).
\end{aligned}
\label{eq:gluon_time_ordering}
\end{equation}
In the Feynman gauge, the free gluon operator is
\begin{equation}
H^\mu{}_\nu=\delta^\mu{}_\nu\nabla_\alpha\nabla^\alpha,
\label{eq:gluon_free_operator_propagator}
\end{equation}
where we have used the vanishing Ricci tensor of Minkowski spacetime. The propagator is a bitensor, with its two Lorentz indices associated with the spacetime points $x$ and $x'$, respectively. It satisfies the Green equation
\begin{equation}
H^\mu{}_\alpha D^{ab\,\alpha\nu'}(x,x')=i\delta^{ab}g^{\mu\nu'}(x,x')\frac{\delta^{(4)}(x-x')}{\sqrt{-g(x)}}.
\label{eq:gluon_green_equation}
\end{equation}
Here, $g^{\mu\nu'}(x,x')$ denotes the parallel propagator with its second index raised. When $x'=x$, it reduces to the inverse metric $g^{\mu\nu}(x)$.

As shown in Sec.~\ref{subsec2.3}, the gluon operator becomes diagonal when the transverse components are expressed in the circular basis
\begin{equation}
A^{(\pm)}=\frac{A^{\hat\rho}\pm iA^{\hat\varphi}}{\sqrt{2}}.
\label{eq:gluon_circular_components_propagator}
\end{equation}
In this basis, the operator takes the form
\begin{equation}
H=\operatorname{diag}(H_0,H_+,H_-,H_0),
\label{eq:gluon_diagonal_operator_propagator}
\end{equation}
where
\begin{equation}
H_0=\Box_\Omega, \qquad
H_\pm=\Box_\Omega+\frac{1}{\rho^2}\mp\frac{2i}{\rho^2}\partial_\varphi.
\label{eq:gluon_diagonal_operators_propagator}
\end{equation}
The corresponding eigenfunctions are
\begin{equation}
\Phi_\alpha^{(s)}(x)=\frac{1}{(2\pi)^{3/2}}e^{-iEt+ik_z z+il\varphi}J_{l+s}(k_\perp\rho),
\quad s=0,\pm1,
\label{eq:gluon_normalized_eigenfunctions}
\end{equation}
with eigenvalues
\begin{equation}
\lambda_\alpha=-(E+\Omega l)^2+k_z^2+k_\perp^2.
\label{eq:gluon_eigenvalues_propagator}
\end{equation}
The orthogonality and completeness relations of these eigenfunctions are established in
Appendix~\ref{app:orthogonality}.

For each sector, we introduce the Feynman inverse
\begin{equation}
G_s(x,x')=\sum_l\int dE\,dk_z\,k_\perp dk_\perp\,\frac{\Phi_\alpha^{(s)}(x)\Phi_\alpha^{(s)*}(x')}{\lambda_\alpha-i\epsilon},
\label{eq:gluon_scalar_green_functions}
\end{equation}
with $s=0,\pm1$. These functions satisfy
\begin{equation}
H_sG_s(x,x')=\frac{\delta^{(4)}(x-x')}{\sqrt{-g}}.
\label{eq:gluon_scalar_green_equations}
\end{equation}
Although the differential operator is diagonal in the circular basis, the propagator is not diagonal because the metric couples the two circular components. In the ordered basis $(\hat t,+,-,\hat z)$, the metric is
\begin{equation}
\eta_{\mathrm{circ}}=
\begin{pmatrix}
1&0&0&0\\
0&0&-1&0\\
0&-1&0&0\\
0&0&0&-1
\end{pmatrix}.
\label{eq:gluon_circular_metric}
\end{equation}
The gluon propagator in this basis is therefore
\begin{equation}
\mathcal D^{ab}(x,x')=i\delta^{ab}
\begin{pmatrix}
G_0&0&0&0\\
0&0&-G_+&0\\
0&-G_-&0&0\\
0&0&0&-G_0
\end{pmatrix}.
\label{eq:gluon_propagator_circular}
\end{equation}
For completeness, substituting the spectral representations of the Green functions $G_s(x,x')$ into Eq.~(\ref{eq:gluon_propagator_circular}), we obtain the full gluon propagator explicitly in the circular basis $(\hat t,+,-,\hat z)$

\begin{widetext}
\begin{equation}
\begin{aligned}
D^{ab}(x,x')={}&-\frac{i\delta^{ab}}{(2\pi)^3}\sum_{l=-\infty}^{\infty}\int_{-\infty}^{\infty}dE\int_{-\infty}^{\infty}dk_z\int_0^\infty dk_\perp\,k_\perp \frac{e^{-iE(t-t')+ik_z(z-z')+il(\varphi-\varphi')}}{(E+\Omega l)^2-k_z^2-k_\perp^2+i\epsilon} \\
&\times
\begin{pmatrix}
J_l(k_\perp\rho)J_l(k_\perp\rho')&0&0&0\\
0&0&-J_{l+1}(k_\perp\rho)J_{l+1}(k_\perp\rho')&0\\
0&-J_{l-1}(k_\perp\rho)J_{l-1}(k_\perp\rho')&0&0\\
0&0&0&-J_l(k_\perp\rho)J_l(k_\perp\rho')
\end{pmatrix}.
\end{aligned}
\label{eq:gluon_propagator_explicit}
\end{equation}
\end{widetext}
Applying the diagonal operator to this expression reproduces the Green equation with the metric
$\eta_{\mathrm{circ}}$.

We now return to the real orthonormal basis. Using
\begin{equation}
A^{\hat\rho}=\frac{A^{(+)}+A^{(-)}}{\sqrt{2}},
\qquad A^{\hat\varphi}=\frac{A^{(+)}-A^{(-)}}{i\sqrt{2}},
\label{eq:gluon_inverse_circular_transformation}
\end{equation}
and defining
\begin{equation}
G_\Sigma=\frac{G_++G_-}{2},
\qquad
G_\Delta=\frac{G_+-G_-}{2},
\label{eq:gluon_green_combinations}
\end{equation}
we obtain
\begin{equation}
D^{ab\,\hat\alpha\hat\beta}(x,x')=\delta^{ab}
\begin{pmatrix}
iG_0&0&0&0\\
0&-iG_\Sigma&G_\Delta&0\\
0&-G_\Delta&-iG_\Sigma&0\\
0&0&0&-iG_0
\end{pmatrix}.
\label{eq:gluon_propagator_orthonormal}
\end{equation}
To make the spectral structure of the gluon propagator explicit, we substitute the eigenfunction expansions of $G_0$, $G_+$, and $G_-$ into Eq.~(\ref{eq:gluon_propagator_orthonormal}). Introducing the radial functions
\begin{align}
\mathcal{J}_0(\rho,\rho') &=J_l(k_\perp\rho)J_l(k_\perp\rho'),\nonumber \\
\mathcal{J}_\pm(\rho,\rho') &=J_{l\pm1}(k_\perp\rho)J_{l\pm1}(k_\perp\rho'),
\end{align}
the gluon propagator in the real orthonormal basis ($\hat t,\hat\rho,\hat\varphi,\hat z$) takes the form

\begin{widetext}
\begin{equation}
\begin{aligned}
D^{ab\,\hat\alpha\hat\beta'}(x,x')={}&-\frac{i\delta^{ab}}{(2\pi)^3}\sum_{l=-\infty}^{\infty}\int_{-\infty}^{\infty}dE\int_{-\infty}^{\infty}dk_z\int_0^\infty dk_\perp\,k_\perp \\
&\times\frac{e^{-iE(t-t')+ik_z(z-z')+il(\varphi-\varphi')}}{(E+\Omega l)^2-k_z^2-k_\perp^2+i\epsilon}
\begin{pmatrix}
\mathcal{J}_0 & 0 & 0 & 0\\
0 & -\dfrac{\mathcal{J}_++\mathcal{J}_-}{2} & -\dfrac{i(\mathcal{J}_+-\mathcal{J}_-)}{2} & 0\\[2mm]
0 & \dfrac{i(\mathcal{J}_+-\mathcal{J}_-)}{2} & -\dfrac{\mathcal{J}_++\mathcal{J}_-}{2} & 0\\
0 & 0 & 0 & -\mathcal{J}_0
\end{pmatrix}.
\end{aligned}
\label{eq:gluon_propagator_orthonormal_explicit}
\end{equation}
\end{widetext}
This expression displays the complete spectral dependence of the free gluon propagator in rotating cylindrical coordinates. The temporal and longitudinal components involve Bessel functions of order ($l$), whereas the transverse components contain the combinations of Bessel functions with orders ($l+1$) and ($l-1$). The off-diagonal transverse terms arise from the transformation between the circular and real orthonormal bases and encode the angular dependence associated with the vector character of the gluon field. 

The functions $G_+$ and $G_-$ also satisfy
\begin{equation}
G_s(x',x)=G_{-s}(x,x'), \qquad s=0,\pm1.
\label{eq:gluon_green_exchange}
\end{equation}
This follows from the changes of integration variables $E\to-E$, $k_z\to-k_z$, and $l\to-l$, together with $J_{-n}(z)=(-1)^nJ_n(z)$. Consequently,
\begin{equation}
G_\Sigma(x',x)=G_\Sigma(x,x'),
\quad
G_\Delta(x',x)=-G_\Delta(x,x').
\label{eq:gluon_green_symmetry}
\end{equation}
These relations ensure that the reconstructed propagator has the symmetry expected for time-ordered bosonic fields,
\begin{equation}
D^{ab\,\hat\alpha\hat\beta}(x,x')=D^{ba\,\hat\beta\hat\alpha}(x',x).
\label{eq:gluon_propagator_exchange_symmetry}
\end{equation}

To express the propagator in the original rotating coordinates, we use the transformation
\begin{equation}
A^\mu=M^\mu{}_{\hat\alpha}A^{\hat\alpha},
\qquad
M^\mu{}_{\hat\alpha}(x)=
\begin{pmatrix}
1&0&0&0\\
0&1&0&0\\
-\Omega&0&1/\rho&0\\
0&0&0&1
\end{pmatrix},
\label{eq:gluon_coordinate_transformation_matrix}
\end{equation}
where the coordinate indices are ordered as $(t,\rho,\varphi,z)$. The propagator becomes
\begin{equation}
D^{ab\,\mu\nu}(x,x')=M^\mu{}_{\hat\alpha}(x)M^\nu{}_{\hat\beta}(x')D^{ab\,\hat\alpha\hat\beta}(x,x').
\label{eq:gluon_propagator_coordinates}
\end{equation}
The transformation matrices are evaluated at their respective spacetime points, as required by the tensorial nature of the propagator.

We can now establish the relation with the propagator in inertial coordinates. Introducing the inertial frequency $\omega_I=E+\Omega l$ and using $\varphi_I=\varphi+\Omega t$, the scalar Green functions can be written as
\begin{equation}
\begin{aligned}
G_s&(x_I,x_I')=-\frac{1}{(2\pi)^3}\sum_{l=-\infty}^{\infty}\int d\omega_I\,dk_z\,k_\perp dk_\perp \\
&\times \frac{e^{-i\omega_I\Delta t_I+ik_z\Delta z_I+il\Delta\varphi_I}J_{l+s}(k_\perp\rho)J_{l+s}(k_\perp\rho')}{\omega_I^2-k_z^2-k_\perp^2+i\epsilon}.
\end{aligned}
\label{eq:gluon_green_inertial}
\end{equation}
Since the sum extends over all integer values of $l$, the change of index $n=l+s$ gives
\begin{align}
\sum_l e^{il\Delta\varphi_I}&J_{l+s}(k_\perp\rho)J_{l+s}(k_\perp\rho')=e^{-is\Delta\varphi_I} \nonumber \\
&\times \sum_n e^{in\Delta\varphi_I}J_n(k_\perp\rho)J_n(k_\perp\rho').
\label{eq:gluon_bessel_index_shift}
\end{align}
It follows that
\begin{equation}
G_\pm(x_I,x_I')=e^{\mp i\Delta\varphi_I}G_0(x_I,x_I').
\label{eq:gluon_green_phase_relation}
\end{equation}
Therefore,
\begin{equation}
G_\Sigma=\cos(\Delta\varphi_I)G_0, \qquad G_\Delta=-i\sin(\Delta\varphi_I)G_0.
\label{eq:gluon_green_trigonometric}
\end{equation}
Substituting these expressions into Eq.~(\ref{eq:gluon_propagator_orthonormal}), we obtain
\begin{equation}
D^{ab\,\hat\alpha\hat\beta}(x_I,x_I')=i\delta^{ab}G_0
\begin{pmatrix}
1&0&0&0\\
0&-\cos\Delta\varphi_I&-\sin\Delta\varphi_I&0\\
0&\sin\Delta\varphi_I&-\cos\Delta\varphi_I&0\\
0&0&0&-1
\end{pmatrix}.
\label{eq:gluon_propagator_inertial_cylindrical}
\end{equation}
The angular dependence reflects the different orientations of the cylindrical bases at $x_I$ and $x_I'$. Transforming both vector indices to Cartesian inertial coordinates removes these angular factors. To recover the Cartesian representation, we introduce the rotation matrix
\begin{equation}
\mathsf R(\varphi_I)=\begin{pmatrix}
\cos\varphi_I & -\sin\varphi_I\\
\sin\varphi_I & \cos\varphi_I
\end{pmatrix}.
\end{equation}
The transformation of both transverse indices gives
\begin{equation}
\mathsf R(\varphi_I)\mathsf R(-\Delta\varphi_I)\mathsf R^T(\varphi_I')=\mathbb I_2,
\end{equation}
where $\Delta\varphi_I=\varphi_I-\varphi_I'$. Consequently, the angular dependence of the transverse tensor structure disappears in the Cartesian basis, leaving the Minkowski metric.

Finally, the Bessel addition theorem gives
\begin{equation}
\sum_{l=-\infty}^{\infty}e^{il\Delta\varphi_I}J_l(k_\perp\rho)J_l(k_\perp\rho')=J_0(k_\perp d_\perp),
\label{eq:gluon_bessel_addition}
\end{equation}
where
\begin{equation}
d_\perp^2=\rho^2+\rho'^2-2\rho\rho'\cos(\Delta\varphi_I).
\label{eq:gluon_transverse_distance}
\end{equation}
Using the integral representation of $J_0$, we recover
\begin{equation}
G_0(x_I,x_I')=-\int\frac{d^4k}{(2\pi)^4}\frac{e^{-ik\cdot(x_I-x_I')}}{k^2+i\epsilon}.
\label{eq:gluon_scalar_minkowski_green}
\end{equation}
The complete gluon propagator in Cartesian inertial coordinates is therefore
\begin{equation}
D^{ab\,\mu\nu}(x_I,x_I')=-i\delta^{ab}\eta^{\mu\nu}\int\frac{d^4k}{(2\pi)^4}\frac{e^{-ik\cdot(x_I-x_I')}
}{k^2+i\epsilon}.
\label{eq:gluon_minkowski_propagator}
\end{equation}
Here, the Lorentz indices $\mu$ and $\nu$ refer to components in the Cartesian inertial frame. This is the standard free gluon propagator in the Feynman gauge. Its recovery confirms that the angular structure of the propagator in the rotating cylindrical basis originates from the coordinate transformation and the position dependence of the vector basis.

As in the ghost sector, the absence of boundaries allows the explicit angular-velocity dependence to be removed by transforming to inertial coordinates. For a system confined within a cylinder of finite radius, the allowed eigenfunctions and transverse momenta must be determined from appropriate boundary conditions. We address this problem in the following subsection.

\subsection{Boundary conditions in a finite cylinder \label{subsec3.3}}

The boundary conditions for gauge fields in a finite domain are not unique. Two geometrically distinct choices are the relative and absolute boundary conditions, which constrain the tangential and normal components of the gauge potential, respectively~\cite{Esposito:1994bv,Vassilevich:2003xt}. Their physical interpretation can be understood in terms of the chromoelectric and chromomagnetic fields measured in the local comoving frame with the cylindrical boundary.

Relative boundary conditions imply
\begin{equation}
\mathbf E_\parallel^a\big|_R=0, \qquad \mathbf B_\perp^a\big|_R=0,
\end{equation}
which are analogous to the electromagnetic boundary conditions of a perfect electric conductor. Conversely, absolute boundary conditions correspond to
\begin{equation}
\mathbf E_\perp^a\big|_R=0,
\qquad
\mathbf B_\parallel^a\big|_R=0,
\end{equation}
and are analogous to those of a perfect magnetic conductor. These analogies characterize the geometry of the field constraints and do not assume a microscopic color-conducting material at the boundary. In the following, we adopt relative boundary conditions for the gluon field and Dirichlet conditions for the Faddeev--Popov ghosts.

We now consider the fields confined to a cylinder of radius $R$, with $0\leq\rho\leq R$ and $|\Omega| R<1$. The latter condition ensures that the rotating Killing vector remains timelike throughout the domain. Unlike the unbounded case, the boundary restricts the allowed transverse momenta and requires boundary conditions compatible with the gauge-fixed theory.

We adopt relative boundary conditions for the gluon field, together with Dirichlet conditions for the Faddeev--Popov ghosts:
\begin{align}
A_t^a\big|_R &= A_\varphi^a\big|_R= A_z^a\big|_R=0, \nonumber\\
\left.\nabla_\mu A^{a\mu}\right|_R &=0, \qquad c^a\big|_R=\bar c^a\big|_R=0.
\label{eq:relative_bc}
\end{align}
Using the orthonormal components defined in Eq.~(\ref{eq:gluon_orthonormal_components}), these conditions become
\begin{align}
& A^{a\hat t}\big|_R=A^{a\hat\varphi}\big|_R=A^{a\hat z}\big|_R=0, \nonumber\\
&\left.\left(\partial_\rho+\frac{1}{R}\right)A^{a\hat\rho}\right|_R=0.
\label{eq:orthonormal_bc}
\end{align}
The second condition follows from the covariant divergence, since the tangential components vanish at the boundary.

For the ghost field, the regular radial eigenfunctions are $J_l(k_\perp\rho)$. Dirichlet conditions therefore give
\begin{equation}
J_l(k_\perp R)=0, \qquad k_{\perp,ln}^{(D)}=\frac{j_{|l|,n}}{R},
\label{eq:ghost_discrete_spectrum}
\end{equation}
where $j_{|l|,n}$ denotes the $n$th positive zero of $J_{|l|}$. The normalized radial functions are
\begin{equation}
u_{ln}^{(D)}(\rho)=\frac{\sqrt{2}\,J_l(k_{\perp,ln}^{(D)}\rho)}{R\,|J_{l+1}(j_{|l|,n})|}.
\label{eq:ghost_radial_modes}
\end{equation}
They satisfy 
\begin{equation}
    \int_0^R\rho\,d\rho\,u_{ln}^{(D)}u_{lm}^{(D)}=\delta_{nm}.
\end{equation}
Consequently, the ghost propagator becomes
\begin{equation}
\begin{aligned}
D_{\mathrm{gh},R}^{ab}(x,x')={}&\frac{i\delta^{ab}}{(2\pi)^3}\sum_{l=-\infty}^{\infty}\sum_{n=1}^{\infty}\int_{-\infty}^{\infty}dE\int_{-\infty}^{\infty}dk_z\\
&\times\frac{e^{-iE\Delta t+ik_z\Delta z+il\Delta\varphi}\,u_{ln}^{(D)}(\rho)u_{ln}^{(D)}(\rho')}{(E+\Omega l)^2-k_z^2-(k_{\perp,ln}^{(D)})^2+i\epsilon}.
\end{aligned}
\label{eq:ghost_finite_cylinder}
\end{equation}

For the transverse gluon components, the differential operators remain diagonal in the circular basis, whereas the boundary conditions couple their amplitudes. We consider
\begin{equation}
A^{(\pm)}=a_\pm J_{l\pm1}(k_\perp\rho)e^{-iEt+ik_z z+il\varphi}.
\end{equation}
Using the second relation in Eq.~(\ref{eq:gluon_inverse_circular_transformation}), the boundary condition $A^{\hat\varphi}|_R=0$ yields $a_+J_{l+1}(x)-a_-J_{l-1}(x)=0$, with $x=k_\perp R$. The radial boundary condition
$(\partial_\rho+1/\rho)A^{\hat\rho}|_R=0$ provides a second equation relating the coefficients $a_+$ and $a_-$. To express this radial boundary condition in terms of the Bessel functions, we use the recurrence relations
\begin{align}
J_{l+1}'(x)&=J_l(x)-\frac{l+1}{x}J_{l+1}(x), \nonumber\\
J_{l-1}'(x)&=-J_l(x)+\frac{l-1}{x}J_{l-1}(x).
\end{align}
Applying these identities to the radial boundary condition and combining the result with $A^{\hat\varphi}|_R=0$, we obtain
\begin{equation}
\begin{pmatrix}
J_{l+1}(x) & -J_{l-1}(x)\\
J_l(x)-\dfrac{l}{x}J_{l+1}(x) &
-J_l(x)+\dfrac{l}{x}J_{l-1}(x)
\end{pmatrix}
\begin{pmatrix}
a_+\\
a_-
\end{pmatrix}
=0.
\label{eq:gluon_boundary_matrix}
\end{equation}
The first row of Eq.~(\ref{eq:gluon_boundary_matrix}) allows us to simplify the second row, yielding
\begin{equation}
(a_+-a_-)J_l(x)=0.
\end{equation}
This relation leads to two families of transverse modes. The first corresponds to $J_l(x)=0$. Using the Bessel recurrence relation $J_{l-1}(x)+J_{l+1}(x)=2lJ_l(x)/x$, we obtain $J_{l-1}(x)=-J_{l+1}(x)$. Substitution into the first row of Eq.~(\ref{eq:gluon_boundary_matrix}) then gives $a_-=-a_+$. This defines the Dirichlet family. The second family corresponds to $a_-=a_+$. In this case, the first row of Eq.~(\ref{eq:gluon_boundary_matrix}) requires $J_{l+1}(x)-J_{l-1}(x)=0$. Since $J_l'(x)=[J_{l-1}(x)-J_{l+1}(x)]/2$, this condition is equivalent to $J_l'(x)=0$, which defines the Neumann family.

A nontrivial solution requires the determinant to vanish. Using $J_{l-1}(x)-J_{l+1}(x)=2J_l'(x)$, we obtain
\begin{equation}
\det M=J_l(x)\left[J_{l-1}(x)-J_{l+1}(x)\right]=2J_l(x)J_l'(x)=0.
\label{eq:gluon_boundary_determinant}
\end{equation}
Consequently, the positive transverse eigenvalues belong to two families,
\begin{equation}
k_{\perp,ln}^{(D)}=\frac{j_{|l|,n}}{R},
\qquad
k_{\perp,ln}^{(N)}=\frac{j'_{|l|,n}}{R},
\label{eq:gluon_discrete_spectra}
\end{equation}
where $j_{|l|,n}$ and $j'_{|l|,n}$ denote the positive zeros of $J_{|l|}$ and $J_{|l|}'$, respectively. 

As shown above, the two transverse families correspond to distinct combinations of the circular components. For the Dirichlet family, $a_-=-a_+$, whereas for the Neumann family, $a_-=a_+$. We introduce the normalized radial eigenfunctions
\begin{equation}
u_{ln}^{(X)}(\rho)=N_{ln}^{(X)}J_l(k_{ln}^{(X)}\rho), \qquad X=D,N,
\end{equation}
with
\begin{align}
N_{ln}^{(D)}&=\frac{\sqrt{2}}{R|J_{l+1}(j_{|l|,n})|}, \nonumber\\
N_{ln}^{(N)}&=\frac{\sqrt{2}}{R|J_l(j'_{|l|,n})|\sqrt{1-l^2/(j'_{|l|,n})^2}}.
\end{align}
Here and below, the transverse eigenvalues are strictly positive. The normalized vector eigenfunctions in the orthonormal basis $(\hat t,\hat\rho,\hat\varphi,\hat z)$ are
\begin{align}
\mathcal T_{ln}^{\hat\alpha}&=(u_{ln}^{(D)},0,0,0), \nonumber\\
\mathcal Z_{ln}^{\hat\alpha}&=(0,0,0,u_{ln}^{(D)}), \nonumber\\
\mathcal P_{ln}^{\hat\alpha}&=\frac{1}{k_{ln}^{(D)}}\left(0,\partial_\rho u_{ln}^{(D)},\frac{il}{\rho}u_{ln}^{(D)},0\right), \nonumber\\
\mathcal Q_{ln}^{\hat\alpha}&=\frac{1}{k_{ln}^{(N)}}\left(0,-\frac{il}{\rho}u_{ln}^{(N)},\partial_\rho u_{ln}^{(N)},0\right).
\label{eq:cylinder_vector_modes}
\end{align}
The functions in Eq.~(\ref{eq:cylinder_vector_modes}) are normalized radial polarization profiles. The corresponding spatial eigenfunctions are obtained by multiplying them by $e^{il\varphi}/\sqrt{2\pi}$. Their orthonormality is defined with respect to the positive definite inner product on the transverse disk,
\begin{equation}
\langle\mathcal U,\mathcal V\rangle=\int_0^{2\pi}d\varphi\int_0^R\rho\,d\rho\,\sum_{\hat\alpha}\mathcal U^{\hat\alpha *}(\rho,\varphi)\mathcal V^{\hat\alpha}(\rho,\varphi).
\end{equation}
The sum over spatial components is understood for the transverse polarizations. In particular, the orthogonality between the gradient-type and curl-type modes follows from integration by parts, since the resulting boundary term vanishes by the Dirichlet condition. Their completeness follows from the Helmholtz decomposition of vector fields on the disk, together with the completeness of the scalar Dirichlet and Neumann eigenfunctions. The constant Neumann zero mode does not generate an independent vector mode, because its transverse gradient vanishes. Furthermore, the simply connected geometry of the disk admits no nontrivial harmonic vector modes satisfying the relative boundary conditions. Consequently, the four polarization families provide a complete basis for the spectral representation of the gluon propagator.

The gluon propagator can then be expressed as a sum over these vector eigenfunctions. Defining
\begin{equation}
\mathcal K_{ln}^{(X)}=(E+\Omega l)^2-k_z^2-(k_{ln}^{(X)})^2+i\epsilon,
\end{equation}
we obtain
\begin{widetext}
\begin{equation}
\begin{aligned}
D_R^{ab,\hat\alpha\hat\beta'}(x,x')={}&-\frac{i\delta^{ab}}{(2\pi)^3}\sum_{l=-\infty}^{\infty}\int_{-\infty}^{\infty}dE\int_{-\infty}^{\infty}dk_z\,e^{-iE\Delta t+ik_z\Delta z+il\Delta\varphi} \\
&\times\left[\sum_{n=1}^{\infty}\frac{\mathcal T_{ln}^{\hat\alpha}(\rho)\mathcal T_{ln}^{\hat\beta *}(\rho')-\mathcal Z_{ln}^{\hat\alpha}(\rho)\mathcal Z_{ln}^{\hat\beta *}(\rho')-\mathcal P_{ln}^{\hat\alpha}(\rho)\mathcal P_{ln}^{\hat\beta *}(\rho')}{\mathcal K_{ln}^{(D)}}-\sum_{n=1}^{\infty}\frac{\mathcal Q_{ln}^{\hat\alpha}(\rho)
\mathcal Q_{ln}^{\hat\beta *}(\rho')}{\mathcal K_{ln}^{(N)}}\right].
\end{aligned}
\label{eq:gluon_finite_cylinder}
\end{equation}
For comparison with the unbounded result, the gluon propagator can also be written explicitly in matrix form. Introducing $x_X=k_{ln}^{(X)}\rho$ and $x_X'=k_{ln}^{(X)}\rho'$, with $X=D,N$, we obtain
\begin{equation}
\begin{aligned}
D_R^{ab,\hat\alpha\hat\beta'}(x,x')={}&-\frac{i\delta^{ab}}{(2\pi)^3}\sum_{l=-\infty}^{\infty}\int_{-\infty}^{\infty}dE\int_{-\infty}^{\infty}dk_z\,e^{-iE\Delta t+ik_z\Delta z+il\Delta\varphi} \\
&\times\Bigg\{
\sum_{n=1}^{\infty}
\frac{(N_{ln}^{(D)})^2}{\mathcal K_{ln}^{(D)}}
\begin{pmatrix}
J_l(x_D)J_l(x_D') & 0 & 0 & 0\\
0 & -J_l'(x_D)J_l'(x_D')
& \dfrac{il}{x_D'}J_l'(x_D)J_l(x_D') & 0\\
0 & -\dfrac{il}{x_D}J_l(x_D)J_l'(x_D')
& -\dfrac{l^2}{x_Dx_D'}J_l(x_D)J_l(x_D') & 0\\
0 & 0 & 0 & -J_l(x_D)J_l(x_D')
\end{pmatrix}
\\
&\qquad+
\sum_{n=1}^{\infty}
\frac{(N_{ln}^{(N)})^2}{\mathcal K_{ln}^{(N)}}
\begin{pmatrix}
0 & 0 & 0 & 0\\
0 & -\dfrac{l^2}{x_Nx_N'}J_l(x_N)J_l(x_N')
& \dfrac{il}{x_N}J_l(x_N)J_l'(x_N') & 0\\
0 & -\dfrac{il}{x_N'}J_l'(x_N)J_l(x_N')
& -J_l'(x_N)J_l'(x_N') & 0\\
0 & 0 & 0 & 0
\end{pmatrix}
\Bigg\}.
\end{aligned}
\label{eq:gluon_cylinder_explicit}
\end{equation}
\end{widetext}

The spectral representations obtained above provide the free gluon and Faddeev--Popov ghost propagators in a finite cylindrical domain. Their construction incorporates the discrete transverse spectra determined by the boundary conditions, while preserving the characteristic energy shift $E+\Omega l$ associated with rotating coordinates. Both propagators satisfy their corresponding Green equations and the imposed boundary conditions.

In the formal limit $R\to\infty$, the discrete radial expansions recover the continuous spectral representations derived for the unbounded domain. However, this limit cannot be taken at fixed nonzero angular velocity while preserving the causality condition $|\Omega|R<1$. A physically admissible sequence of cylinders with increasing radius necessarily requires $\Omega\to0$.

\section{Conclusions \label{sec4}}

In this work, we have constructed the free gluon and Faddeev--Popov ghost propagators in rigidly rotating cylindrical coordinates using an eigenfunction approach. Starting from the quadratic gauge-fixed QCD action in the Feynman gauge, we derived the corresponding kinetic operators and determined their spectral representations. For the gluon field, the introduction of a circular polarization basis allowed us to diagonalize the vector wave operator and identify the polarization-dependent shifts in the orders of the radial Bessel functions. This construction provides an explicit treatment of the tensor structure of the gluon propagator while preserving the angular-momentum dependence of the underlying field modes.

In the absence of radial boundaries, we showed that the off-diagonal components of the gluon propagator in the cylindrical orthonormal basis arise from the relative orientation of the local basis vectors at different spacetime points. By transforming to inertial Cartesian coordinates, we recovered the standard Minkowski propagator. This result establishes that the additional angular structure appearing in the cylindrical representation is geometrical in origin and does not, by itself, represent a dynamical modification of the free gluon propagator induced by rotation.

We extended the spectral construction to a cylinder of finite radius, imposing relative boundary conditions on the gluon field and Dirichlet conditions on the ghost fields. These conditions lead to discrete transverse spectra and organize the gluon eigenfunctions into families associated with Dirichlet and Neumann eigenvalues. Using the corresponding normalized modes, we obtained explicit spectral representations of both propagators, including the full tensor structure of the gluon propagator. The finite-radius formulation therefore incorporates the effects
of the boundary through the allowed eigenfunctions and eigenvalues, while retaining the angular and polarization
structure required by the vector nature of the gluon field.

Beyond providing a consistent spectral description of free gauge and ghost fields in rotating cylindrical geometries, our results establish a framework for perturbative calculations in QCD under rotation. In particular, the explicit mode dependence of the propagators provides the ingredients for constructing correlation functions and radiative corrections while accounting for angular momentum, polarization, and finite-radius boundary conditions. The simultaneous treatment of gluon and ghost propagators within the same gauge-fixed formulation is especially relevant for future studies of gluonic self-energies and other loop corrections. Such applications will allow the interplay between rotation, confinement to a finite spatial domain, and the dynamics of strongly interacting fields to be investigated beyond the free-field level.

\begin{acknowledgments}
Support for this work was received by the Secretaría de Ciencia, Humanidades, Tecnología e Innovación Grant No. CBF-2025-G-1718. R. Z. acknowledges support from the following grant ANID/FONDECYT/Regular No. 1241436. B. G. C. acknowledges the financial support of a fellowship granted by Secretaría de Ciencia, Humanidades, Tecnología e Innovación as part of the Sistema Nacional de Posgrados.
\end{acknowledgments}

\appendix

\section{Orthogonality and completeness of the eigenfunction basis \label{app:orthogonality}}

In this appendix, we establish the orthogonality and completeness relations of the eigenfunctions used to construct the free gluon and ghost propagators. We consider the unbounded radial domain $0\leq\rho<\infty$, for which the transverse momentum is continuous.

The eigenfunctions of the diagonal operators $H_s$, with $s=0,\pm1$, are given by
\begin{equation}
\Phi_\alpha^{(s)}(x)=\mathcal{N}_s e^{-iEt+ik_z z+il\varphi}J_{l+s}(k_\perp\rho),
\label{eq:app_eigenfunctions}
\end{equation}
where $\alpha=(E,k_z,k_\perp,l)$, with $E,k_z\in\mathbb{R}$, $k_\perp\geq0$, and $l\in\mathbb{Z}$. The sector $s=0$ corresponds to the ghost and the temporal and longitudinal gluon components, whereas $s=\pm1$ corresponds to the circular transverse gluon components.

The invariant integration measure in rotating cylindrical coordinates is
\begin{equation}
d^4x\,\sqrt{-g}=dt\,dz\,d\varphi\,\rho\,d\rho,
\label{eq:app_invariant_measure}
\end{equation}
where $\sqrt{-g}=\rho$. We use this measure to define the distributional inner product
\begin{equation}
\left\langle \Phi_\alpha^{(s)},\Phi_{\alpha'}^{(s)}\right\rangle=\int d^4x\,\sqrt{-g}\,\Phi_\alpha^{(s)*}(x)\Phi_{\alpha'}^{(s)}(x).
\label{eq:app_inner_product}
\end{equation}

\subsection{Orthogonality}

Substituting Eq.~(\ref{eq:app_eigenfunctions}) into Eq.~(\ref{eq:app_inner_product}), we obtain
\begin{equation}
\begin{aligned}
\left\langle\Phi_\alpha^{(s)},\Phi_{\alpha'}^{(s)}\right\rangle={}&|\mathcal{N}_s|^2\int_{-\infty}^{\infty}dt\,e^{i(E-E')t} \\
&\times\int_{-\infty}^{\infty}dz\,e^{-i(k_z-k_z')z} \\
&\times \int_0^{2\pi}d\varphi\,e^{i(l'-l)\varphi} \\
&\times \int_0^\infty\rho\,d\rho\, J_{l+s}(k_\perp\rho) J_{l'+s}(k_\perp'\rho).
\end{aligned}
\label{eq:app_orthogonality_integral}
\end{equation}
The temporal and longitudinal integrations give
\begin{equation}
\begin{aligned}
\int_{-\infty}^{\infty}dt\,e^{i(E-E')t}&=2\pi\delta(E-E'), \\
\int_{-\infty}^{\infty}dz\,e^{-i(k_z-k_z')z} &=2\pi\delta(k_z-k_z'),
\end{aligned}
\label{eq:app_fourier_orthogonality}
\end{equation}
while the angular integration yields
\begin{equation}
\int_0^{2\pi}d\varphi\,e^{i(l'-l)\varphi}=2\pi\delta_{ll'}.
\label{eq:app_angular_orthogonality}
\end{equation}
The Kronecker delta enforces $l'=l$, so the radial integral contains Bessel functions of the same order. We can therefore use the Hankel orthogonality relation
\begin{equation}
\int_0^\infty\rho\,d\rho\,J_n(k_\perp\rho)J_n(k_\perp'\rho)=\frac{\delta(k_\perp-k_\perp')}{k_\perp},
\label{eq:app_bessel_orthogonality}
\end{equation}
valid in the distributional sense for integer $n$ and $k_\perp,k_\perp'>0$.

Combining these results, we find
\begin{equation}
\begin{aligned}
\left\langle\Phi_\alpha^{(s)},\Phi_{\alpha'}^{(s)}\right\rangle={}&(2\pi)^3|\mathcal{N}_s|^2\delta(E-E')\delta(k_z-k_z') \\
&\times \delta_{ll'}\frac{\delta(k_\perp-k_\perp')}{k_\perp}.
\end{aligned}
\label{eq:app_orthogonality_result}
\end{equation}
Choosing
\begin{equation}
\mathcal{N}_s=\frac{1}{(2\pi)^{3/2}},
\qquad s=0,\pm1,
\label{eq:app_normalization}
\end{equation}
we obtain the normalized orthogonality relation
\begin{equation}
\begin{aligned}
\int d^4x\,\sqrt{-g}\,\Phi_\alpha^{(s)*}(x)\Phi_{\alpha'}^{(s)}(x)={}&\delta(E-E')\delta(k_z-k_z') \\
&\times\delta_{ll'}\frac{\delta(k_\perp-k_\perp')}{k_\perp}.
\end{aligned}
\label{eq:app_normalized_orthogonality}
\end{equation}
The normalization is independent of both the angular velocity $\Omega$ and the label $s$.

\subsection{Completeness}

We now consider the mode expansion
\begin{equation}
\begin{aligned}
    \mathcal{C}_s(x,x')={}&\sum_{l=-\infty}^{\infty}\int_{-\infty}^{\infty}dE\int_{-\infty}^{\infty}dk_z\int_0^\infty k_\perp\,dk_\perp\\
    &\times\Phi_\alpha^{(s)}(x)\Phi_\alpha^{(s)*}(x').
\end{aligned}
\label{eq:app_completeness_expansion}
\end{equation}
Substituting the normalized eigenfunctions gives
\begin{equation}
\begin{aligned}
\mathcal{C}_s(x,x')={}&\frac{1}{(2\pi)^3}\int_{-\infty}^{\infty}dE\,e^{-iE(t-t')} \\
&\times \int_{-\infty}^{\infty}dk_z\, e^{ik_z(z-z')} \\
&\times \sum_{l=-\infty}^{\infty} e^{il(\varphi-\varphi')} \\
&\times \int_0^\infty k_\perp\,dk_\perp\, J_{l+s}(k_\perp\rho) J_{l+s}(k_\perp\rho').
\end{aligned}
\label{eq:app_completeness_integral}
\end{equation}
The Fourier integrals yield
\begin{equation}
\begin{aligned}
\int_{-\infty}^{\infty}dE\,e^{-iE(t-t')}&=2\pi\delta(t-t'),\\
\int_{-\infty}^{\infty}dk_z\,e^{ik_z(z-z')}&=2\pi\delta(z-z').
\end{aligned}
\label{eq:app_fourier_completeness}
\end{equation}
For the radial integral, we use the Hankel completeness relation
\begin{equation}
\int_0^\infty k_\perp\,dk_\perp\,J_n(k_\perp\rho)J_n(k_\perp\rho')=\frac{\delta(\rho-\rho')}{\rho},
\label{eq:app_bessel_completeness}
\end{equation}
which holds for any integer order $n$. In particular, it applies to $n=l+s$ for all three sectors.

The remaining angular sum gives
\begin{equation}
\sum_{l=-\infty}^{\infty}e^{il(\varphi-\varphi')}=2\pi\delta_{2\pi}(\varphi-\varphi'),
\label{eq:app_angular_completeness}
\end{equation}
where $\delta_{2\pi}$ denotes the periodic Dirac delta on the angular interval.

Combining these identities, we obtain
\begin{equation}
\mathcal{C}_s(x,x')=\frac{1}{\rho}\delta(t-t')\delta(z-z')\delta(\rho-\rho')\delta_{2\pi}(\varphi-\varphi').
\label{eq:app_completeness_result}
\end{equation}
Since $\sqrt{-g}=\rho$, the completeness relation takes the covariant form
\begin{equation}
\sum_{l=-\infty}^{\infty} \int dE\,dk_z\,k_\perp dk_\perp\, \Phi_\alpha^{(s)}(x) \Phi_\alpha^{(s)*}(x')=\frac{\delta^{(4)}(x-x')}{\sqrt{-g}},
\label{eq:app_normalized_completeness}
\end{equation}
for $s=0,\pm1$.

Thus, the eigenfunctions of each diagonal operator form a complete generalized basis with the spectral measure
\begin{equation}
d\mu(\alpha)=\sum_{l=-\infty}^{\infty}dE\,dk_z\,k_\perp dk_\perp.
\label{eq:app_spectral_measure}
\end{equation}
These relations provide the normalization and completeness properties required for the eigenfunction expansion of the free gluon and ghost propagators. They apply to the unbounded radial domain; the corresponding relations for a finite cylinder must be constructed using the eigenfunctions and
boundary conditions of that problem.

\bibliography{references}

\end{document}